\documentclass[runningheads]{svmult}

\usepackage{makeidx}   
\usepackage{graphicx}  
\usepackage{subeqnar}  
\usepackage{multicol}  
\usepackage{physprbb}  
\makeindex             

\begin{document}
\title*{The local Larmor clock,
Partial Densities of States, and Mesoscopic Physics}
%
%
%
%
\titlerunning{Larmor clock, Density of States}
%
\author{Markus B\"uttiker}
\authorrunning{Markus B\"uttiker}
%
%
\institute{D\'epartement de physique th\'eorique,
Universit\'e de Gen\`eve, 24 Quai Ernest-Ansermet \\ 
CH-1211 Gen\`eve, Switzerland}
\maketitle              

\begin{abstract}
Starting from the Larmor clock we introduce a hierarchy 
of density of states. At the bottom of this hierarchy are the 
partial densities of states which represent the contribution 
to the local density of states if both the incident and 
the out-going scattering channel is prescribed. 
We discuss the role 
of the partial densities of states in a number of 
electrical conduction problems in phase coherent mesoscopic systems: 
The partial densities of states play a prominent role 
in measurements with a scanning tunneling microscope 
on multiprobe conductors in the presence of current flow.
The partial densities of states determine the degree of dephasing 
generated by a weakly coupled voltage probe. 
We show that the 
partial densities of states determine the frequency-dependent 
response of mesoscopic conductors in the presence of slowly oscillating 
voltages applied to the contacts of the sample.
We introduce the off-diagonal elements of the partial density of 
states matrix to describe fluctuation processes.  These examples 
demonstrate that the  
analysis of the Larmor clock has a wide range of applications.

\end{abstract}

\section{Introduction}
\label{section1}
\begin{figure}
\begin{center}
\includegraphics[scale=0.9]{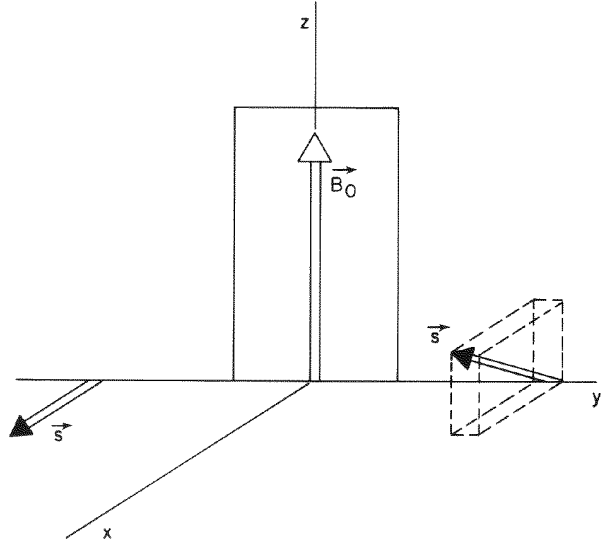}
\end{center}
\caption{Spin polarized carriers incident on a barrier 
subject to a weak magnetic field $B_0$. 
The transmitted carriers exhibit both 
a spin precession and and a spin rotation. After \protect\cite{LARMOR} . }
\label{clock}
\end{figure}

The Larmor clock is one of the most widely discussed
approaches to determine the time-scales of tunneling processes. 
The essential idea\cite{BAZ,RYB,LARMOR} of the Larmor clock is that the motion 
of the spin polarization in a narrow region of magnetic field 
can be exploited to provide information on the time carriers spend 
in this region. It is assumed that incident carriers are spin
polarized and that they impinge on a region to which a small magnetic 
field is applied perpendicular to the spin polarization of the incident 
carriers (see Fig. \ref{clock}). The spin polarization of the 
transmitted and reflected carriers
can then be compared with the polarization of the incident carriers. 
Dividing the angle between the polarization of the exciting carriers
and that of the incident carrier by the Larmor 
frequency $\omega_L$ gives a time. 
Originally, only spin precession (the movement of the polarization
in the plane perpendicular to the magnetic field) was considered. 
However, Ref. \cite{LARMOR} pointed out, that especially if 
we deal with regions in which only evanescent waves exist (tunneling
problems) the polarization executes not only a precession but also
a rotation into the direction of the magnetic field. In fact in the 
presence of a tunneling barrier, the spin {\it rotation}, is the dominant 
effect. Ref. \cite{LARMOR} considered a rectangular barrier
and considered a magnetic field of the same spatial extend 
as the barrier. In the local version of the Larmor clock, 
introduced by Leavens and Aers\cite{LEAE1}, we consider an arbitrary region 
in which the magnetic field is non-vanishing and investigate again 
the direction of the spin polarization and rotation of the transmitted and 
reflected carriers. The magnetic field might be non-vanishing
in a small region localized inside the barrier or in a small 
region outside barrier on the side on which carriers are incident or 
on the far side of the barrier. We mention already here, that 
the response of the carriers is highly non-local: Even carriers which are 
reflected are affected by a magnetic field that is non-vanishing only
on the far side of the barrier where naively we would expect only 
transmitted carriers\cite{LEAE2}. 
In this work we use the local 
Larmor clock to derive a set of local density of states 
states \cite{BU1,BTP,BTP1,GASP,GRAM} 
which we call {\it partial densities of states} 
which are related to {\it spin precession} and in terms of 
{\it sensitivities} which are related to {\it spin rotation}. The
partial densities of states, below abbreviated as PDOS, are useful
to understand a number of transport problems: 
the transmission probability from a tunneling microscope
tip into a multiterminal mesoscopic conductor \cite{GRAM} can be expressed
in terms of PDOS, the absorption of carriers by an optical potential
(a potential with a small imaginary component),
inelastic scattering and dephasing caused by a weak coupling voltage probe, 
and the low frequency transport in mesoscopic conductors.  

The partial densities of states are determined by functional 
derivatives of the scattering matrix \cite{BU1,BTP,GASP,GRAM}. 
Only in certain limited situations can the density of states
be expressed in terms of energy derivatives. Expressions for the 
density of states in terms of energy derivatives of the scattering matrix 
are familiar \cite{dash,avba}. 
In the discussion of characteristic times the distinction 
between time-scales found from energy derivatives  (like 
the Wigner-Smith phase delay) and time scales 
found from derivatives with respect to the local potential 
(the dwell time) has found 
some recognition. In contrast density of states are almost invariable 
discussed in terms of energy derivatives. Here we emphasize that 
a more precise discussion of density of states also uses derivatives 
with respect to the (local) potential and not energy derivatives. 
It is the dwell time (or sums of relevant dwell times) 
which are related to 
the density of states \cite{ianna1,ianna2}. 
The use of energy derivatives always signals that approximations 
are involved. 

The interpretation of the Larmor clock remains a subject of 
discussion. Ref. \cite{LARMOR} considered the total 
rotation angle dived by the Larmor precession frequency to be the 
relevant time. This interpretation brings the Larmor clock 
into agreement with the time-scales obtained by considering 
tunneling through a barrier with an oscillating potential \cite{Land}. 
Subsequent works have argued that the precession angle 
and the rotation angle dived by the Larmor
precession frequency separately should be viewed as 
time scales \cite{SOKOL,GOR}. 
The difficulty with such an interpretation is not only that 
one has two scales characterizing the same process, but 
the times defined in such a way are also not necessarily positive. 
Since we aim at characterizing a duration, that is a definite 
draw back. 
The two time scales can be combined into a complex time, with the 
real part referring to the precession time and 
the imaginary part to the time obtained from rotation. 
Like negative times, complex times are not part of the commonly
accepted notions of time. Steinberg argues that the clock 
presents only a "weak measurement" and that therefore complex times 
are permitted \cite{Stein}. 
In quantum mechanics the questions "how much 
time has the transmitted particle spent in a given interval" 
is problematic since being in an "interval" and "to be transmitted"
corresponds to noncommuting operators \cite{muga}.  
Reasonably, we can only speak of a time duration, if it is real 
and positive.

A comparison of the Larmor clock with the closely related 
linear ac-response of an electrical conductor shows immediately 
the ambiguity of the clock: in the ac-response of a conductor 
which is predominantly capacitive (tunneling limit) 
the voltage leads the current whereas for a highly transmissive 
conductor the response is inductive and the current leads the voltage.
 
The language 
used here implies similarly an extension of the usual notion 
of density of states. 
At the bottom of the hierarchy of density of states 
which we discuss are 
the {\it partial densities of states} (PDOS) which represents
the contribution to the local density of states 
if we prescribe both the incident and the out-going channel. 
It turns out that certain partial densities of states are not positive. 
(These are of course just the PDOS that 
correspond to negative precession times). 
Thus the discussion presented below does not 
solve the interpretational questions related to the Larmor clock. 
Nevertheless as we will show even such negative PDOS 
are physically relevant. Using the
partial densities of states, either by summing over the 
out-going channels (or by summing over the incident channels) 
we obtain the {\it injectivity} of a contact into a point within the sample 
or the {\it emissivity} of a point within the sample into a contact. 
Both injectivities and emissivities are positive and in the language 
of tunneling times correspond to local dwell times for which either 
the incident channel (or the outgoing channel but not both) 
are prescribed. Finally, if we take the sum of all the injectivities 
or the sum of all the emissivities we obtain the local density of states. 

\section{The scattering matrix and the local Larmor clock}
\label{section2}

We start by considering a one-dimensional scattering 
problem \cite{LARMOR,LEAE1,LEAE2}. 
We consider particles moving along the $y$-axis in a potential 
$V(y)$. The potential is arbitrary, except that 
asymptotically, for large negative and large positive values of $y$
it is assumed to be flat. We adopt here the language from mesoscopic 
transport discussions and call the region of large 
negative $y$ the contact $1$ (the left contact) 
and the region of large positive $y$ contact $2$ (the right contact). 
We assume that the quantum mechanical evolution
is described by the Schr\"odinger equation. A particle with energy
$E$ has for large negative $y$ a wave vector $k_1(E)$ and a velocity
$v_1(E)$ and for large positive values of $y$ 
a wave vector $k_2(E)$ and a velocity
$v_2(E)$. We are interested in scattering states. 
A particle incident from the left 
is for large negative values of $y$ described by a scattering state 
\begin{equation}
\psi_1(E,y) = e^{ik_{1}y} + S_{11} \, e^{-ik_{1}y}
\label{s11}
\end{equation}
and for large positive values of $y$ is described by a transmitted wave 
\begin{equation}
\psi_1(E,y) = (\frac{v_{1}}{v_{2}})^{1/2} S_{21}\,  e^{ik_{2}y} .
\label{s21}
\end{equation}
Similarly, a particle incident from the right 
is for large positive values of $y$ described by a scattering state 
\begin{equation}
\psi_2(E,y) = e^{-ik_{2}y} + S_{22} \, e^{-ik_{2}y}
\label{s22}
\end{equation}
and for large negative values of $y$ is described by a transmitted wave 
\begin{equation}
\psi_2(E,y) = (\frac{v_{2}}{v_{1}})^{1/2} S_{12} \, e^{ik_{2}y} .
\label{s12}
\end{equation}
Here the amplitudes $S_{\alpha\beta}$ 
determine the elements of the $2*2$-scattering matrix 
of the problem. Each scattering matrix element is 
a function of the energy $E$ of the incident carrier and 
is a functional of the potential $V(y)$. To express this dependence
we write  $S_{\alpha\beta}(E,V(y))$. 
Conservation of the probability current requires that this matrix 
is unitary and in the absence of a magnetic field time-reversal invariance
implies that it is also symmetric. 

Next we now consider a weak magnetic field applied to a small region.
The magnetic field shall point into the $z$-direction.
For simplicity we consider the case where the magnetic
field is constant in a small interval $[y,y+dy]$
and takes there the value $B$. 
We consider only
the effect of the Zeeman energy. Thus the motion of the particle 
remains one-dimensional and is as in the absence of a magnetic field
confined to the $y$-axis. In the set-up for the Larmor clock, 
we consider particles with a spin. The spin of the incident particles
(in the asymptotic regions) is polarized along the $x$-axis. 
For spin 1/2 particles 
the wave functions are now spinors with two components 
$\psi_{+}(y,E)$ and $\psi_{-}(y,E)$. 
Carriers incident from the left, have a spinor with components given by 
$\psi_{+}(y,E) = \psi_{-}(y,E) = ({1/\sqrt 2}) exp(ik_1y)$.
The Zeeman energy which is generated by the local magnetic field
is diagonal in the spin up and spin down components.
Consequently, in the interval $[y,y+dy]$  
for a particle with spin up, 
the energy is reduced by 
$\hbar \omega_L/2 $ with $\omega_L/2 =g \mu B/\hbar$
and for a particle with spin down the potential is increased 
by $\hbar \omega_L/2 $. Thus with the magnetic field 
switched on particles with spin up travel in a potential 
$V(y) - dV (y)$ and particles with spin down travel in 
potential $V(y) + dV (y)$. Here $V(y)$ is the potential 
in the absence of the magnetic field and $dV(y)$
is the potential generated by the magnetic field. 
Thus  $dV(y)$ vanishes everywhere, except in the interval
$[y,y+dy]$ where it takes the value 
$dV(y) = \hbar \omega_L/2 = g \mu B$.  

We can evaluate the polarization of the transmitted particles and the 
reflected particles if we can determine the scattering matrix for 
spin up and spin down particles in the potential generated by the magnetic 
field. Thus we need the scattering matrices 
$S^{\pm}_{\alpha\beta}(E,V(y) \mp dV(y))$
where $S^{+}$ is the scattering matrix for spin up carriers and 
$S^{-}$ is the scattering matrix for spin down carriers. 
Since the potential variation generated by the magnetic field is small
we can expand these matrices away from the scattering matrix for 
the unperturbed potential. Thus we find to first order in the magnetic field 
for the scattering matrices 
\begin{eqnarray}\label{fs1}
S^{\pm}_{\alpha\beta}(E,V(y)) \mp dV(y)) = 
S_{\alpha\beta}(E,V(y)) 
\mp 
[dS_{\alpha\beta}(E,V(y))/dV(y)] dV(y) dy + .....
\end{eqnarray}
The variation of the scattering matrix due to the magnetic 
field is proportional to the derivative of the scattering matrix
with respect to the local potential at the location where 
the magnetic field is non-vanishing. 
More generally, we can consider a magnetic field 
which varies along the $y$-axis (but always points along
the $z$-axis). This leads to a potential $\delta V(y)$
determined by the local magnetic field. 
The variation of the scattering matrix is then determined by
a functional derivative 
$[\delta s_{\alpha\beta}(E,V(y))/\delta V(y)]$
of the scattering matrix with regard 
to the local potential,  
\begin{eqnarray}\label{fs2}
S^{\pm}_{\alpha\beta}(E,V(y)) \mp \delta V(y)) = 
S_{\alpha\beta}(E,V(y)) \nonumber\\
\mp \int dy^{\prime} 
[\delta s_{\alpha\beta}(E,V(y^{\prime}))/\delta V(y^{\prime})]
\delta  V(y^{\prime}) + ..... .
\end{eqnarray}
We emphasize that even though this equation looks quite simple, the 
evaluation of a functional derivative of a scattering matrix, 
while not difficult for a one-dimensional problem, 
can still be a laborious calculation. 

Let us now find the precession and rotation angles of the polarization
of the transmitted and reflected carriers. The normalized spinor of the 
transmitted particles which determines the spin orientation of the transmitted 
a particles has the components 
\begin{equation}
\psi_{1+}(E,y) = \frac{S^{+}_{21}}{(|S^{+}_{21}|^{2}+|S^{-}_{21}|^{2})^{1/2}}.
\label{spt1}
\end{equation}
\begin{equation}
\psi_{1-}(E,y) = \frac{S^{-}_{21}}{(|S^{+}_{21}|^{2}+|S^{-}_{21}|^{2})^{1/2}}.
\label{spt2}
\end{equation} 
First consider the 
polarization in the y direction. It is found by evaluating the expectation
value of the Pauli spin matrix $\sigma_y$, 
\begin{equation}
<s_{y}>_{21} = \frac{\hbar}{2} <\psi_{1} |\sigma_{y}|\psi_{1}>
= - i \frac{\hbar}{2} \frac{S^{+\dagger}_{21}S^{-}_{21}-
S^{+}_{21}S^{-\dagger}_{21}}
{(|S^{+}_{21}|^{2}+|S^{-}_{21}|^{2})} .
\label{sy1}
\end{equation}  
Here the indices $21$ indicate that we consider transmission 
from left ($1$) to right ($2$) and evaluate the the spin in the 
transmitted beam. 
We need the spin polarization only to first order
in the applied magnetic field. 
Using Eq. (\ref{fs2}) we find 
\begin{equation}
<s_{y}>_{21} = \frac{h}{T} \nu(2,y,1) \omega_L dy
\label{spin1}
\end{equation}
where $T = |S_{21}|^{2}$ 
is the transmission probability 
in the absence of the magnetic field and 
\begin{equation}
\nu(2,y,1) = -\frac{1}{4\pi i}
\left( S_{21}^{\dagger}
\frac{\delta S_{21}}{\partial V(y)} - 
\frac{\delta S_{21}^{\dagger}}{\delta V(y)}
S_{21}\right)\;\;
\label{lpdos1}
\end{equation}
is the partial density of states at $y$ of carriers
which emanate from contact $1$
(the asymptotic region for large negative 
$y$) and eventually in the future reach contact $2$.  
Since initially the spin polarization was in the $x$-direction 
$<s_{y}>_{21}$ directly determines the angle of precession of the carriers
in the $x-y$-plane. Thus by dividing $<s_{y}>_{21}$ by the Larmor precession
frequency we can formally introduce a quantity with the dimension of 
time, which we call $\tau_{y}(2,y,1)$ and which is given by 
$\tau_{y}(2,y,1) = ({h}/{T}) \nu(2,y,1) dy . $
Here the index $y$ indicates that we deal with a time-scale obtained 
from the $y$-component of the spin polarization. We can now proceed to evaluate
also the $y$-component of the spin polarization of the carriers which 
are reflected and can proceed to evaluate the 
$y$-component of the spin polarization 
of the carriers that are in the past incident from contact $2$ 
(large positive $y$) and in the future will be transmitted into 
contact $1$ (large negative  $1$) or will be reflected back into contact $1$. 
We can summarize the results in the following manner: 
There are a total of four spin polarizations to be considered, 
each of them determined by a partial density of states  
\begin{equation}
\nu(\alpha, y, \beta ) = -\frac{1}{4\pi i}
\left( S_{\alpha \beta}^{\dagger}
\frac{\delta S_{\alpha \beta}}{\delta V(y)} - 
\frac{\delta S_{\alpha \beta}^{\dagger}}{\delta V(y)}
S_{\alpha \beta}\right)\;\;
\label{lpdos2}
\end{equation} 
of carriers that are incident at contact $\beta = 1, 2$ 
and eventually in the future are transmitted or reflected 
into contact $\alpha = 1, 2$.  
Formally, the time scales related to precession in the local magnetic 
field at $y$ can be  introduced which are related to the partial 
densities of states via, 
$\tau_{y}(\alpha, y, \beta ) = ({h}/{|S_{\alpha \beta}|^{2}}) 
\nu(\alpha, y, \beta) dy$
where $|S_{\alpha \beta}|^{2}$ is the transmission probability $T$
if $\alpha$ and $\beta$ are not equal and is the reflection 
probability $R$ if $\alpha$ and $\beta$ are equal. 
Thus with {\it each} element of the scattering
matrix we can associate a partial density of states.
Later we discuss the properties of the partial densities 
of states in more detail. 

Next we consider the spin polarization in the $z$-direction. 
For the carriers incident in contact $1$ and transmitted into contact $2$ 
we find that the $z$-component of the transmitted carriers is determined by 
\begin{equation}
<s_{z}>_{21} = \frac{\hbar}{2} <\psi_{1} |\sigma_{z}|\psi_{1}>
= \frac{\hbar}{2} \frac{|S^{+}_{21}|^{2} - |S^{-}_{21}|^{2}}
{(|S^{+}_{21}|^{2}+|S^{-}_{21}|^{2})}.
\label{sz1}
\end{equation} 
Using Eq. (\ref{fs2}) we find 
\begin{equation}
<s_{z}>_{1} = \frac{h}{T} \eta(2,y,1) \omega_L dy . 
\label{sz2}
\end{equation} 
where we call 
\begin{equation}
\eta(2, y, 1) = - \frac{1}{4\pi}
\left( S_{21}^{\dagger}
\frac{\delta S_{21}}{\delta U(y)} + 
\frac{\delta S_{21}^{\dagger}}{\delta U(y)}
S_{21}\right)\;\;
\label{lsens}
\end{equation} 
the {\it sensitivity} of the scattering problem.  
Since the spin polarization of the incident particles 
was originally along the $x$-direction only a small $z$-component
of the incident particle determines a spin {\it rotation} angle. 
We can formally introduce a time scale 
$\tau_{z}(2,y,1)$ associated with spin rotation 
which is given by 
$\tau_{z}(2,y,1) = ({h}/{T}) \eta(2,y,1) dy $. 
Again we can ask about the $z$-polarization of reflected particles
and can ask about the $z$-polarization of particles incident from 
the right. The results are summarized by attributing each scattering 
matrix element $S_{\alpha \beta}$ a sensitivity 
\begin{equation}
\eta(\alpha, y, \beta) = - \frac{1}{4\pi}
\left( S_{\alpha \beta}^{\dagger}
\frac{\delta S_{\alpha \beta}}{\delta U(y)} + 
\frac{\delta S_{\alpha \beta}^{\dagger}}{\delta U(y)}
S_{\alpha \beta}\right)\;\;
\end{equation} 
which determine the time-scales 
$\tau_{z}(\alpha, y, \beta)= ({h}/{|S_{\alpha \beta}|^{2}}) 
\eta(\alpha, y, \beta) dy .$ 
Finally we can determine the spin polarization in the $x$-direction. 
This component is reduced from its initial value 
both because of spin precession 
in the $x-y$-plane and because of the rotation of spins into
the $z$-direction. 
Since we have at every space point 
\begin{equation}
<s_{x}>^{2} + <s_{y}>^{2} + <s_{z}>^{2} = \hbar^{2}/4 ,
\end{equation}
it follows immediately that the time scale $\tau_{x}$
is related to the two time-scales introduced above by 
\begin{equation}
\tau_{x} = (\tau_{y}^{2} + \tau_{z}^{2})^{1/2} . 
\end{equation}
Using the expressions for $\tau_{y}$ and $\tau_{z}$ 
given above, we find for the time-scales $\tau_{x}$
the following expressions 
\begin{equation}
\tau_{x}(\alpha, y, \beta) = \frac{h}{|S_{\alpha \beta}|^{2}}
\left(\frac{\delta S_{\alpha \beta}}{\delta V(y)}  
 \frac{\delta S_{\alpha \beta}^{\dagger}}{\delta V(y)}\right)^{1/2}.\;\;
\end{equation}
We reemphasize that neither the partial densities of states
nor the sensitivities are in general positive.  
In contrast, $\tau_{x}(\alpha, y, \beta)$ is positive
for all elements of the scattering matrix.  

\section{Absorption and Emission of Particles: Injectivities
and Emissivities}

Before discussing the partial densities of states in more detail it is 
of interest to investigate the absorption of particles in a small 
scattering region \cite{MB90}. 
We assume that in a narrow interval $[y,y+dy]$
there exists a non-vanishing absorption rate $\Gamma$.
Thus the potential $V(y)$ is equal to $V_{0}(y)-i\hbar \Gamma$ 
in the interval $[y,y+dy]$ and is equal to $V_0(y)$
out-side this interval. 
To solve the scattering problem 
we need to find the scattering matrix $S^{\Gamma}_{\alpha\beta}(E,V(y))$
in the presence of this complex potential $V(y)$. 
Of interest here is, as in Ref. \cite{MB90}, the limit of small 
absorption. The case of strong absorption 
is also of interest but thus has been used only to discuss 
global properties and not the local quantities of interest 
here \cite{rama,been}. 
For a small absorption rate we can expand the scattering matrix 
$S^{\Gamma}_{\alpha\beta}(E,V(y))$ in powers of the absorption rate away
from the scattering problem in the original real potential $V_{0}$. 
We obtain 
\begin{eqnarray}\label{g1}
& & S^{\Gamma}_{\alpha\beta}(E,V(y)) = 
S_{\alpha\beta}(E,V_{0}(y)) \nonumber\\
& & + i \hbar [\delta S_{\alpha\beta}(E,V(y))/\delta V(y)]|_{V(y) =V_{0}(y)} 
\Gamma dy + .....
\end{eqnarray}
We note that the adjoint scattering matrix has to be evaluated in the 
potential $V^{*} (y)$ and hence 
\begin{eqnarray}\label{g2}
& & S^{\Gamma\dagger}_{\alpha\beta}(E,V(y)) = 
S^{\dagger}_{\alpha\beta}(E,V_{0}(y)) \nonumber\\
& & - i \hbar 
[\delta S^{\dagger}_{\alpha\beta}(E,V(y))/\delta V(y)]|_{V(y) =V_{0}(y)} 
\Gamma dy + ....
\end{eqnarray} 
With these results it easy to show that the 
transmission and reflection probabilities in the presence of 
a small absorption in the interval $[y,y+dy]$ are 
\begin{equation}
|S^{\Gamma}_{\alpha\beta}(E,V(y))|^{2} = 
|S_{\alpha\beta}(E,V_{0}(y))|^{2} 
(1- \Gamma \nu(\alpha,y,\beta) dy )
\label{g3}
\end{equation}
where $|S_{\alpha\beta}(E,V_{0}(y))|^{2}$
is the transmission probability $T$ of the scattering problem 
without absorption if $\alpha$ and $\beta$ are different
and is the reflection probability $R$ of the scattering problem 
without absorption if $\alpha = \beta$. The incident current
$j_{in}$, must be equal to the sum of
the transmitted current $j_{T}$, the reflected current $j_{R}$
and the absorbed current $j_{\Gamma}$, 
\begin{equation} 
j_{in} = j_{T} + j_{R} + j_{\Gamma} .
\label{g4}
\end{equation}
Using Eq. (\ref{g1}) and taking into 
account that the incident flux is normalized to $1$, we find for carriers 
incident from the left or right $\beta =1, 2$ an absorbed flux
given by 
\begin{equation} 
j_{\Gamma} (y,\beta) = \Gamma \nu(y,\beta) dy,
\label{g5}
\end{equation}
where $\nu(y,\beta)$ is called the {\it injectivity}
of contact $\beta$ into point $y$. The injectivity of the contact
is related to the partial densities of states via 
\begin{equation} 
\nu(y,\beta) = \sum_{\alpha} \nu(\alpha, y, \beta) .
\label{g6}
\end{equation}
In our problem with two contacts the injectivity is just the sum of two 
partial densities of states. 

Another way of determining the absorbed flux proceeds as follows. 
The absorbed flux is proportional to the integrated density 
of particles in the region of absorption (in the interval $[y,y+dy]$).
The density of particles can be found from the scattering 
state $\psi_{\beta}(y)$ given by Eqs. (\ref{s11} -\ref{s12}).
For carriers incident from contact $\beta$ the absorbed flux is thus 
\begin{equation} 
j_{\Gamma}(y,\beta)  = \Gamma \frac{1}{hv_{\beta}}|\psi_{\beta}(y)|^{2} dy .
\label{g7}
\end{equation}
Note that here the density of states $1/hv_{\beta}$ of the 
asymptotic scattering region appears. It normalizes the incident current 
to $1$. 
Thus we have found a wave function representation for the 
injectivity. Comparing Eq. (\ref{g4}) and Eq. (\ref{g3}) gives
\begin{equation} 
\nu(y,\beta) = \frac{1}{hv_{\beta}}|\psi_{\beta}(y)|^{2} .
\label{g8}
\end{equation}
The total local density of states $\nu(y)$ at point $y$ is 
obtained by considering carriers incident from both contacts. 
In terms of wave functions  $\nu(y)$ is for our one-dimensional
problem given by 
\begin{equation} 
\nu(y) = \sum_{\beta} \frac{1}{hv_{\beta}}|\psi_{\beta}(y)|^{2} .
\label{g9}
\end{equation}
Thus the total density of states is also the sum of 
the injectivities from the left and right contacts 
\begin{equation} 
\nu(y) = \sum_{\beta} \nu(y,\beta).
\label{g10}
\end{equation} 
There is now an interesting additional problem to be addressed. 
Instead of a potential which acts as a carrier sink (as an absorber)
we can ask about a potential which acts as a carrier source. Obviously, 
all we have to do to turn our potential into a carrier source is to 
change the sign of the imaginary part of the potential. With a 
a carrier source in the interval $[y,y+dy]$ we should observe
a particle current toward contact $1$ and a particle current toward 
contact $2$. We suppose that carriers are incident both from the left and 
the right and evaluate the currents in the contact regions. 
The total current injected into the sample at $y$ is 
\begin{equation} 
j_{in} (y) = \Gamma \nu(y) dy. 
\label{g11}
\end{equation}
Taking into account that the incident current is normalized to $1$
the current $j_{out}(\beta,y)$  in contact $\beta$ due to a carrier 
source at $y$ is given by 
\begin{equation}
j_{out}(\alpha,y) = 1- \sum_{\beta} |s^{\Gamma}_{\alpha\beta}(E,V(y))|^{2}
\label{g12}
\end{equation}
due to the modification of both the transmission and reflection coefficients.
Using Eq. (\ref{g1}) (with $\Gamma$ replaced by -$\Gamma$)
gives 
\begin{equation}
j_{out}(\alpha,y) = - \Gamma \sum_{\beta}\nu(\alpha,y,\beta) \, dy
= - \Gamma \nu(\alpha,y) \, dy . 
\label{g13}
\end{equation}
The current in contact $\alpha$ is determined by the 
{\it emissivity} $\nu(\alpha,y)$ of the point $y$ into contact $\alpha$. 
Note the reversal of the sequence of arguments in the emissivity 
as compared to the injectivity. Thus the emissivity is like the 
injectivity a sum of two partial densities of states, 
\begin{equation}
\nu(\alpha,y) = \sum_{\beta}\nu(\alpha,y,\beta) . 
\label{g14}
\end{equation}
For a scattering problem in the absence of a magnetic field
the injectivity and emissivity are identical. If there is 
a homogeneous magnetic field present they are related by reciprocity:
the injectivity from contact $\alpha$
into point $y$ is equal to the emissivity of point $y$ into the 
contact $\alpha$ in a magnetic field that has been reversed, 
$\nu_{+B}(y,\alpha) = \nu_{-B}(\alpha,y)$. 

We have thus obtained a hierarchy of density of states:
At the bottom are the partial densities of states 
$\nu(\alpha,y,\beta)$ for which we describe both 
the contact from which the carriers are incident and the 
contact through which the carriers have to exit. 
On the next higher level are the injectivities $\nu(y,\alpha)$
and the emissivities $\nu(\alpha ,y)$. 
For the injectivity we prescribe the contact through which the carrier
enters but the final contact is not prescribed. In the emissivity
we prescribe the contact through which the carrier leaves
but the incident contact is not prescribed. 
Finally, on the highest level is the local density of states $\nu(y)$
for which we prescribe neither the incident contact nor the contact 
through which carriers leaves.

For simple scattering problems (delta-functions, barriers) 
the interested reader can find a derivation and discussion of 
partial densities of states in Refs.  \cite{GASP,GRAM,zhao}.  

Returning to time scales: we have shown that the partial
densities of states are associated with spin precession. 
It is tempting, therefore, to associate them with 
a time duration. However, as can be shown, the partial
densities of states are not necessarily positive.
(The simple example of a resonant double barrier
shows that  
one of the diagonal elements $\nu(\alpha,y,\alpha)$
has a range of energies where it is negative \cite{MB90}).
The injectivities and emissivities are, however, always positive. 
The proof is given by Eq. (\ref{g6}). 
We can associate a local dwell time 
$\tau_D (y,\beta) = \hbar \nu_D (y,\beta)$
with the injectivity which gives the time a carrier 
incident from contact $\beta$ 
spends in the interval $[y,y+dy]$ irrespective 
of whether it is finally reflected or transmitted. 
Similarly, we can associate a dwell time with the emissivity 
which is the time carriers spend in the interval $[y,y+dy]$
irrespective from which contact they entered the scattering region.  
There is little question that the dwell times have the properties
which we associate with the duration of a process: they are real and 
positive. However, as explained they do not characterize 
transmission or reflection processes. 

\section{Potential Perturbations}

Thus far our discussion has focused much on the partial 
densities of states. The sensitivity introduced as a measure 
of the spin rotation in the Larmor clock
has, however, also an immediate direct interpretation.
We have seen that the partial densities of states are obtained in response
to a complex perturbation of the original potential $V(y)$.
The sensitivity comes into play if we consider a {\it real} perturbation 
$\delta V$ of the original potential. Thus if we consider a potential 
which is equal to $V(y) + \delta V$ in the interval 
$[y,y+dy]$ and equal to $V(y)$ elsewhere
the transmission probability $T^{V}$ in the presence of the perturbation
is $T^{V} = T + 4\pi \eta (\alpha, y,\beta) \delta V dy$, 
with $\alpha \ne \beta$. The reflection probability 
is  $R^{V} = R + 4\pi \eta (\alpha, y, \alpha) \delta V dy$.
Since also $T^{V} + R^{V} =1$ we must have 
$\eta (y) \equiv  \eta (\alpha, y,\beta) = - \eta (\alpha, y, \alpha)$. 
For our scattering problem, described by a $2*2$ scattering matrix 
there exists only one independent sensitivity $\eta (y)$. 
In the Larmor clock the sensitivity corresponds to spin rotation
and the fact that there is only one sensitivity follows from the 
conservation of angular momentum: the weak magnetic field cannot
produce a net angular momentum. If carriers in the transmitted beam 
acquire a polarization in the direction of the magnetic field 
then carriers in the reflected beam must have a corresponding 
polarization opposite to the direction of the magnetic field. 
In mesoscopic physics, in electrical transport problems, the sensitivity
plays a role in the discussion of non-linear current-voltage characteristics
and plays a role if we ask about the change of the conductance 
in response to the variation of a gate voltage \cite{br2}. 
Below, we will not further discuss the sensitivity, but we will present 
a number of examples in which the partial densities of states play 
a role.

\section{Generalized Bardeen formulae}

\begin{figure}
\begin{center}
\includegraphics[scale=0.9]{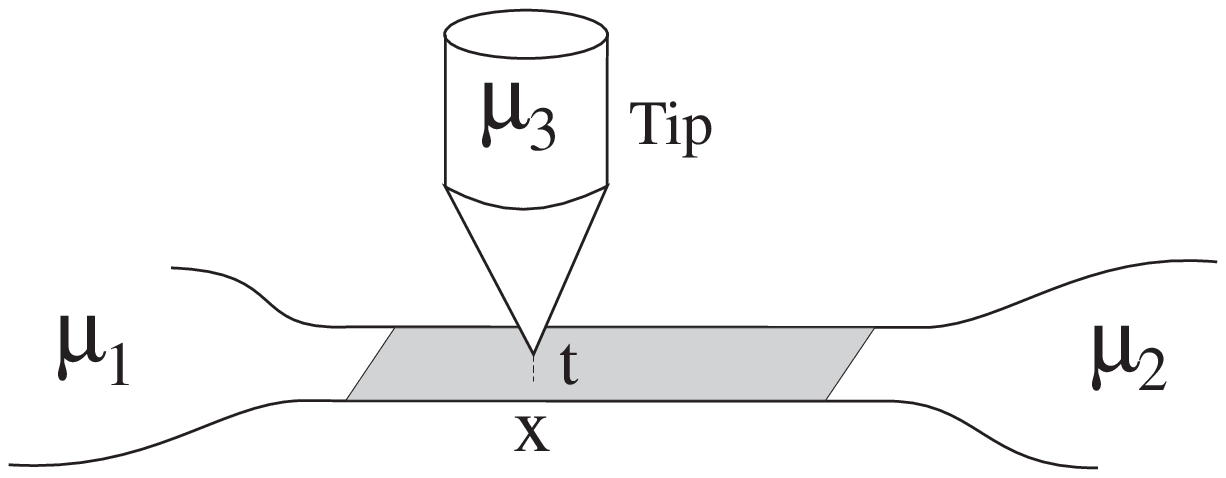}
\end{center}
\caption{Partial density of states measurement with a 
scanning tunneling microscope. 
The tip of an STM couples at a point $x$ with a coupling strength $t$
to the surface of a multi-terminal conductor. 
The contacts of the conductor are
held at potentials $\mu_{\alpha} = eV_\alpha$ 
and the tip at potential $\mu_{3}= eV_{tip}$. After \protect\cite{GRAM}.}
\label{eintip}
\end{figure}

It is well known that with a scanning tunneling microscope (STM)
we can measure the local density of states \cite{STM}. STM measurements
are typically performed in a two terminal geometry, in which
the tip of the microscope represents one contact and the sample 
provides another contact \cite{STM}. Here we 
consider a different geometry. We are interested 
in the transmission 
probability from an STM tip into the contact of a sample with two or more 
contacts as shown in Fig. \ref{eintip}. Thus we deal with a multiterminal 
transmission problem \cite{GRAM}. If we denote the contacts of the sample 
by a Greek letter $\alpha = 1, 2, ..$ and use {\it tip} 
to label the contact of the STM tip, we are interested in the 
tunneling probabilities $T_{\alpha tip}$ from the tip 
into contact $\alpha$ of the sample. In this case the 
STM tip acts as carrier source. Similarly we ask about 
the transmission probability $T_{tip \alpha}$ 
from a sample contact to the tip. In this case the STM tip 
acts as a carrier sink. Earlier work has addressed this problem 
either with the help of scattering matrices, electron wave dividers, 
or by applying the Fermi Golden Rule. Recently, Gramespacher and 
the author \cite{GRAM} have returned to this problem and have derived 
expressions for these transmission probabilities from the
scattering matrix of the full problem (sample plus tip). 
For a tunneling contact with a density of states $\nu_{tip}$
which couples locally at the point 
$x$ with a coupling energy 
$|t|$ these authors found
\begin{equation}
T_{tip,\alpha}=4\pi^2\, \nu_{tip}\, |t|^2\, \nu(x,\alpha)\, ,\label{w1}
\end{equation}
\begin{equation}
T_{\alpha tip}=4\pi^2\, \nu(\alpha,x)\, |t|^2\, \nu_{tip}\, .\label{w2}
\end{equation}
In a multiterminal sample the transmission probability
from a contact $\alpha$
to the STM tip is given by the injectivity of contact $\alpha$ into the 
point $x$ and the transmission probability from the tip 
to the contact $\alpha$ is given by the emissivity of the point $x$
into contact $\alpha$. Eqs. (\ref{w1}) and (\ref{w2})
when multiplied by the unit of conductance $e^{2}/h$ 
are generalized Bardeen conductances for tunneling 
into multiprobe 
conductors. Since the local density of states of the tip
is an even function of magnetic field 
and since the injectivity and emissivity are related by reciprocity
we also have the reciprocity relation $T_{tip,\alpha}(B) = 
T_{\alpha,tip} (-B)$. 

The presence of the tip also affects transmission and reflection 
at the massive contacts of the sample. 
To first order in the coupling energy $|t|^{2}$
these probabilities are 
given by 
\begin{equation}
|S^{tip}_{\alpha\beta}|^{2} = |S_{\alpha\beta}|^{2} - 
4\pi^2\, \nu(\alpha,x,\beta)\, |t|^2\, \nu_{tip}\, .\label{w3}
\end{equation}
The correction to the transmission probabilities $\alpha \ne \beta$
and reflection probabilities $\alpha = \beta$ is determined by 
the partial densities of states, the coupling energy and the 
density of states in the tip. Note that if these probabilities
are placed in a matrix then each row and each column of this 
matrix adds up to the number of quantum channels in the contacts.  

\section{Voltage probe and inelastic scattering}

Consider a two probe conductor much smaller than any inelastic or phase
breaking length. Carrier transport through such a structure can then be said 
to be coherent and its conductance is at zero
temperature given by $G = (e^{2}/h) T$, where $T$ is the 
probability for transmission form one contact to the other. 
How is this result affected by events which break the phase or by 
events which cause inelastic scattering? To investigate this 
question Ref. \cite{MB88} proposes to use 
an additional (third) contact to the sample. 
The third probe acts as a voltage probe which has its potential 
adjusted in such a way that there is no net current  
flowing into this additional probe,
$I_{3} = 0$. The current at the third probe is set to zero by floating
the voltage $\mu_3 = eV_{3}$ at this contact to a 
value for which $I_{3}$ vanishes. 
The third probe acts, therefore, like a {\it voltage probe}.
Even though the total current at the voltage probe vanishes individual 
carriers can enter this probe if they are at the same time replaced 
by carriers emanating from the probe \cite{MB88}. 
Entering and leaving a contact are 
{\it irreversible} processes, since there is no definite phase relationship
between a carrier that enters the contact and a carrier that leaves 
a contact. In a three probe conductor, the relationship 
between currents and voltages is given by $I_{\alpha} = 
\sum_{\beta} G_{\alpha\beta}V_{\beta}$ where the $G_{\alpha\beta}$
are the conductance coefficients. Using the condition $I_{3} =0$
to find the potential $V_{3}$ and eliminating this potential
in the equation for $I_{2}$ or $I_{1}$ gives for the two probe conductance
in the presence of the voltage probe 
\begin{equation}
G = - G_{21} - \frac{G_{23}\, G_{31}}{G_{31}\, +\, G_{32}} .
\label{v1}
\end{equation}
For a very weakly coupled voltage probe we can use Eqs. (\ref{w1} - \ref{w3}).
Taking into account that $G_{\alpha\beta} = - (e^{2}/h) |S_{\alpha\beta}|^{2}$ 
for $\alpha \ne \beta$
we find 
\begin{equation}
G = \frac{e^{2}}{h}\,  \left(T - 4\pi^2\,  |t|^2\,  [\nu(2,x,1) - 
\frac{\nu(2,x)\, \nu(x,1)}{\nu(x)}] \right) . 
\label{v2}
\end{equation} 
Here $\nu(x)$ is the local density of states at the location of the
point at which the voltage probe couples to the conductor. 
Eq. (\ref{v2}) has a simple interpretation \cite{MB88}. 
The first term $T$ is 
the transmission probability of the conductor in the absence of the 
voltage probe. The first term inside the brackets 
proportional to the local partial density of states gives the 
reduction of coherent transmission due to the presence of the 
voltage probe. The second term in the brackets is the incoherent
contribution to transport due to inelastic scattering induced by the 
voltage probe. It is proportional to the injectivity of contact $1$
at point $x$. A fraction $\nu(2,x)/{\nu(x)}$ of the carriers which reach
this point, proportional to its emissivity, are scattered forward 
and, therefore, contribute to transport. Notice the different signs
of these two contributions. The effect of inelastic scattering (or dephasing)
can either enhance transport or diminish transport, depending on 
whether the reduction of coherent transmission (first term) 
or the increase due to incoherent transmission (second term) 
dominates.  

Instead of a voltage probe, we can also use 
an optical potential to simulate inelastic scattering or dephasing. 
However, in order to preserve current, we must use both an absorbing 
optical potential (to take carriers out) and an emitting optical 
potential (to reinsert carriers). The absorbed and re-emitted 
current must again exactly balance each other. 
From Eq. (\ref{g4}) it is seen that the coherent current is again 
diminished by $\Gamma \nu(2,x,1)$, i. e. by the partial density 
of states at point $x$. The total absorbed current is proportional 
to $\Gamma \nu(x,1)$, the injectance of contact $1$ into this point.
As shown in section 3 a carrier emitting optical potential 
at $x$ generates a current $ - \Gamma \nu(1,x)$ in contact $1$
and generates a current $ - \Gamma \nu(2,x) $ in contact $2$.
It produces thus a total current $ - \Gamma \nu(x) $.
In order that the generated and the absorbed current are equal 
we have to normalize the emitting optical potential 
such that it generates a total current proportional to
$\Gamma \nu(x,1)$
(equal to the absorbed current). 
The current at contact $2$ generated by an optical potential
normalized in such a way is thus $-\Gamma \nu(2,x)\nu(x,1)/\nu(x)$. 
The sum of the two contributions, the absorbed current and the 
re-emitted current gives an overall transmission (or conductance)
which is given by Eq. (\ref{v2}) with $4\pi |t|^{2}$ replaced by 
$\Gamma$.   
 
Thus the weakly coupled voltage probe 
(which has current conservation built in)
and a discussion based on optical potentials coupled with a 
current conserving re-insertion of carriers are equivalent \cite{BB}.
There are discussions in the literature which invoke optical
potentials but do not re-insert carriers. Obviously, such 
discussions violate current conservation. 
A recent discussion \cite{JAY}, which compares the voltage probe model and 
the approach via optical potentials, does re-insert carriers 
but does this in an ad hoc manner. In fact Ref. \cite{JAY}
claims that the Onsager symmetry relations are violated
in the optical potential approach. This is an incorrect conclusion
arising from the arbitrary manner in which carriers are re-inserted.   

We conclude this section with a cautionary remark: We 
have found here that the weakly coupled probe voltage probe model 
and the optical potential model are equivalent. But this equivalence 
rests on a particular description of the voltage probe.  There are 
many different models and even in the weak coupling limit our description
of the voltage probe given here is not unique. The claim can only be that 
for sufficiently weak optical absorption and re-insertion 
of carriers there exits {\it one} voltage probe model 
which gives the same answer. Differing weak coupling voltage probes are 
discussed in Refs. \cite{MBOPT}. 

\section{AC Conductance of mesoscopic conductors} 

In this section we discuss as an additional application 
of partial densities of states briefly the ac-conductance of 
mesoscopic systems. We consider a conductor with 
an arbitrary number of 
contacts labeled 
by a Greek index $\alpha = 1, 2, 3...$. 
The problem is to find the relationship 
between the currents $I_{\alpha}(\omega)$ at frequency $\omega$
measured at the contacts of the sample 
in response to a sinusoidal voltage with amplitude $V_{\beta}(\omega)$
applied to contact $\beta$. The relationship between currents 
and voltages is given by a dynamical conductance matrix \cite{BTP}
$G_{\alpha\beta} (\omega)$ such that 
$I_{\alpha}(\omega) = \sum_{\beta}G_{\alpha\beta} (\omega)V_{\beta}(\omega)$.
All electric fields are localized in space. 
The overall charge on the conductor is conserved. 
Consequently, current is also conserved and the currents 
depend only on voltage differences. Current conservation
implies  $\sum_{\alpha}G_{\alpha\beta} = 0$ for each $\beta$.  
In order that only voltage differences matter, the dynamical
conductance matrix has to obey $\sum_{\beta}G_{\alpha\beta} = 0$
for each $\alpha$. We are interested here in the low frequency 
behavior of the conductance and therefore we can expand the conductance in
powers of the frequency \cite{BU1},  
\begin{equation}
G_{\alpha\beta} (\omega) = G^{0}_{\alpha\beta} - i \omega E_{\alpha\beta} 
+ K_{\alpha\beta} \omega^{2} + O(\omega^{3}) .
\label{ac1}
\end{equation} 
Here $G^{0}_{\alpha\beta}$ is the dc-conductance matrix. 
$E_{\alpha\beta}$ is called the {\it emittance} matrix 
and governs the displacement currents. $K_{\alpha\beta}$ 
gives the response to second order in the frequency. 
All matrices $G^{0}_{\alpha\beta}, E_{\alpha\beta}$ and 
$K_{\alpha\beta}$ are real. 

We focus here on the emittance matrix $E_{\alpha\beta}$.
The conservation of the total charge can only be achieved 
by considering the long-range Coulomb interaction. 
Here we describe the long-range Coulomb interaction in a
random phase approach (RPA) in terms of an effective 
interaction. The effective interaction potential $g(x^{\prime},x)$ 
has to be found
by solving a Poisson equation with a non-local screening term.
The effective interaction gives the potential variation 
at point $x^{\prime}$ in response to a variation of the charge 
at point $x$. With the help of the effective interaction 
we find for the emittance matrix \cite{BU1}
\begin{equation}
E_{\alpha\beta} = e^{2} \left[ \int dx \nu(\alpha,x,\beta) 
- \int dx^{\prime} dx
\nu(\alpha,x^{\prime})\, g(x^{\prime},x)\,  \nu(x,\beta) \right] 
\label{ac2}
\end{equation} 
Here the first term, proportional to the integrated 
partial density of states, is the ac-response at low frequencies
which we would have in the absence of interactions. 
The second term has the following simple interpretation:
an ac-voltage applied to contact $\beta$ would (in the absence of
interactions) lead to a charge built up at point $x$ given by 
the injectivity of contact $\beta$. Due to interaction, this charge generates 
at point $x^{\prime}$ a variation in the local potential which then induces 
a current in contact $\alpha$ proportional to the emissivity 
of this point into contact $\alpha$. The effective interaction has the 
property that at an additional charge with a distribution proportional 
to the local density of states gives rise to a spatially uniform potential, 
$\int dx^{\prime} \nu(x^{\prime})g(x^{\prime},x) = 1$ for every $x$. 
This property ensures that the elements of 
each row and each column of the emittance matrix add up 
to zero. In particular if screening is local (over a length scale 
of a Thomas Fermi wave length) we have $g(x^{\prime},x) = 
\delta (x^{\prime} - x) \nu^{-1}(x)$. 
In this limit the close connection between Eq. (\ref{ac2}) 
and Eq. (\ref{w2}) is then obvious. 
\begin{figure}
\begin{center}
\includegraphics[scale=0.7]{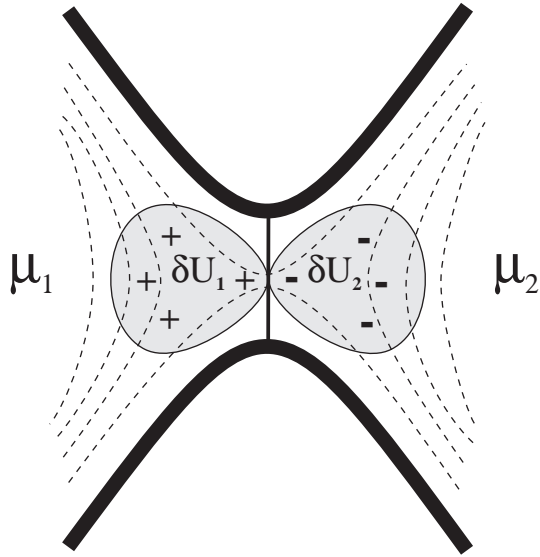}
\end{center}
\vspace{-1cm}
\caption{Charge dipole across a saddle point 
constriction. $\mu_{1} = eV_{1}$ and  $\mu_{2} = eV_{2}$
are the potentials of the contacts, $\delta U_{1}$ and  $\delta U_{2}$
are local potentials. The dashed lines are the equipotential lines 
of the equilibrium potential. 
After \protect\cite{TCMB}.}
\label{potential}
\end{figure}
\section{Transition from Capacitive to Inductive Response} 

The following example \cite{TCMB} provides an instructive application 
of the ac-conduc-
tance formula Eq. (\ref{ac2}). Consider the 
transmission through a narrow opening shown in Fig. \ref{potential}. 
Carrier motion is in two dimension through a potential $U(x,y)$
which has the form a saddle with a height $U_0$. The 
conductance (transmission) through such a narrow opening 
(a quantum point contact) 
has been widely studied and is found to rise step-like \cite{Wees}
as a function of the potential $U_0$ with plateaus 
at values $G = (2e^{2}/h) N$ corresponding to perfect 
transmission of $N$ spin degenerate channels. 
Here we are interested in the emittance $E$ as a function of $U_0$. 
We consider the case that $U_0$ is so large that transmission is 
completely blocked and then lower $U_0$ such that the probability 
of transmission probability $T$ gradually increases from $0$ to $1$. 

\begin{figure}  
\begin{center}
\includegraphics[scale=0.6]{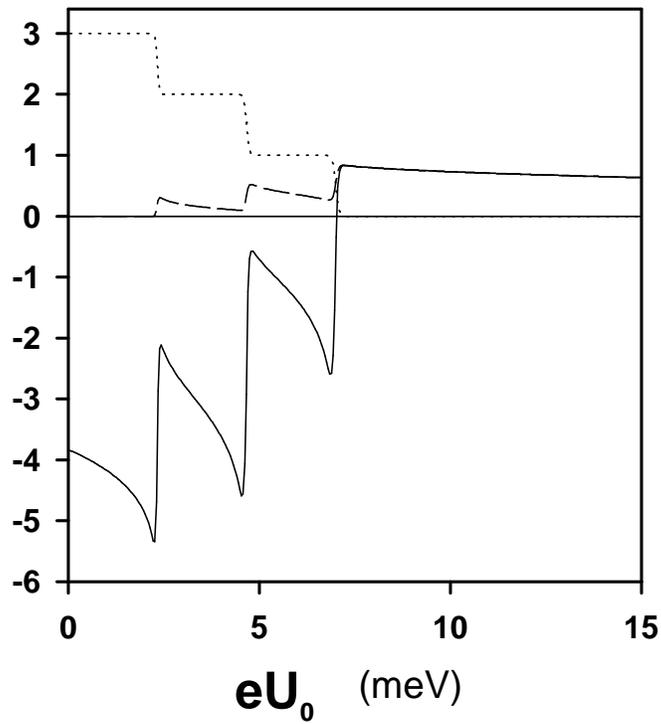}
\end{center}
\vspace{-3.8cm}
\caption{Conductance of the saddle point constriction (dotted line)
in units of $2e^{2}/h$, capacitance and emittance (dashed and full 
curve) as a function of the height of the saddle point potential $U_{0}$.
In the range of voltages shown three quantum channels are opened.
After \protect\cite{TCMB}.\vspace{-0.2cm}}
\label{emit}
\end{figure}

We introduce two regions $\Omega_{1}$ and  $\Omega_{2}$ 
to the left
and the right of the barrier, respectively. 
Instead of the local partial density of states 
we consider the partial density of states 
integrated over the respective volumes $\Omega_{1}$ and $\Omega_{2}$.
Thus we introduce 
$D_{\alpha k \beta} = \int_{\Omega_k} dxdy  \nu(\alpha,x,y,\beta)$. 
We furthermore introduce the total density of states $D$ of the two
regions. We assume that the potential has left-right symmetry 
and consequently the density of states in the regions 
$\Omega_{1}$ and  $\Omega_{2}$ are $D_{1}=D_{2}= D/2 $. 
Ref. \cite{TCMB} evaluates the partial densities of states 
semiclassically. 
We find that carriers incident from contact $1$
and transmitted into contact $2$ give rise to a partial densities
of states in region $1$ given by 
$D_{211}= T D_{1}/2$. To determine $D_{212} $, we note
that in the semiclassical limit considered here, there
are no states in $\Omega _{1}$ associated
with scattering from contact $2$ back to contact $2$, hence it holds
$D_{212} =0$. With similar arguments one finds for the semi-classical PDOS
\begin{equation}
D_{\alpha k \beta} =
D_{k} \left( T/2 + \delta _{\alpha \beta}(R\: \delta _{\alpha
k}-T/2)\:\right) \;\;, \;\;\; {\rm if }\;\; \alpha , \beta \neq 3 \;\;.
\label{lpdosqpc}
\end{equation} 
From Eq. (\ref{lpdosqpc}) we obtain for the emissivity
into contact $1$ from region $1$ and injectivity from contact $1$
into region $1$, 
$D^{e}_{11} = D^{i}_{11}=(1/4)(1+R) D$ and 
and obtain for the emissivity into contact $1$ 
from region $2$ and the injectivity into region $1$
from contact $2$,
$D^{e}_{12} = D^{i}_{12}=(1/4)TD$. 
Instead of the full Poisson equation the 
effective interaction \cite{TCMB,BU1,CURACAO} is 
determined with the help of a geometrical capacitance $C$. 
For a detailed discussion we have to refer the 
reader to Refs. \cite{TCMB,CURACAO}. Due to charge 
conservation we have $E \equiv E_{11} = E_{22} = -E_{12} = -E_{21}$
with 
\begin{eqnarray}
E = (RC_{\mu}- DT^{2}/4) 
\label{MB1}
\end{eqnarray}
Here $C_{\mu}^{-1} = R^{-1} (C^{-1} + (e^{2}D/4)^{-1})$
is the effective capacitance of the contact. It is proportional 
to the reflection probability and proportional to the 
series capacitance of the geometrical capacitance  $C$ 
and the "quantum capacitance" $e^{2}D$.
If the contact is completely
closed we have $R = 1, T = 0$ and the emittance is completely 
determined by the capacitance $C_{\mu}$. If the channel is completely
open we have $R = 0, T = 1$ and the emittance is $E = - D/4$. It is negative
indicating that for a completely open channel the ac-response is now 
not capacitive but inductive. Thus there is a voltage $U_0$ for which 
the emittance vanishes. For a simple saddle point potential 
the behavior of the capacitance and emittance is illustrated 
in Fig. \ref{emit}. The dotted line shows the conductance, 
the dashed line is the capacitance $C_{\mu}$ and the full line 
is the emittance as a function of the saddle point potential $U_{0}$. 
The emittance is capacitive (positive) for a nearly closed contact
and changes sign as the transmission probability increases 
from near $0$ to $1$. The emittance shows additional structure 
associated with the successive opening of further quantum channels. 

A similar transition from capacitive to inductive behavior
is found in the emittance of a mesoscopic wire: Guo and coworkers \cite{Guo} 
investigate the emittance of a wire as a function of impurity 
concentration. A ballistic wire with no impurities has an inductive 
response, a disordered metallic diffusive wire has a capacitive 
response.  Experiments on ac-transport in mesoscopic structures are 
challenging and we can mention here only the work by 
Pieper and Price \cite{pp} 
on the ac-conductance of an Aharonov-Bohm ring 
and the recent work by Desrat et al. \cite{desrat} on 
the low-frequency impedance of quantized Hall conductors.

Our simple example demonstrates 
the difficulty in associating a time with a result 
obtained by analyzing a stationary (or as here) a quasi-stationary 
scattering problem. The emittance divided by the 
conductance quantum $e^{2}/h$ has the dimension of a time. 
(In the non-interacting limit $e^{2}/C << 4/D$ we have \cite{Mikh} 
$E / (e^{2}/h) = (R - T) \tau_{D} /4 $ where $\tau_D$ is the dwell 
time in the two regions $\Omega_1$ and $\Omega_2$).   
But since this "time" changes sign such an interpretation is not 
appropriate. Furthermore, in electrical problems the natural "times" 
are $RC$-times if the low frequency dynamics is capacitive 
or an $R/L$-time if it is inductive.  The fact that we describe 
here a  crossover from capacitive to an inductive-like behavior 
demonstrates that neither of these two time-scales can adequately 
describe the dynamics. 
\section{Partial Density of States Matrix}

Thus far the main aim of our discussion has been to illustrate 
the concept of partial densities of states with a number of 
simple examples. We now would like to point to some important 
extensions of this concept. 

Quantum mechanics is a theory of probability amplitudes. 
We can thus suppose that our initial state 
is a scattering experiment which is described 
by a superposition $\Psi (y) = a_{\beta} \Psi_{\beta}(y)
+ a_{\gamma}\Psi_{\gamma}(y)$. Here $\Psi_{\beta}$ and $\Psi_{\gamma}$
are two scattering states describing particles incident in 
channel $\beta$ {\it and} in channel $\gamma$ with amplitudes 
$a_\beta$ and $a_\gamma$. 
Often such superpositions are 
eliminated since we can assume that 
each incident amplitude carries its own phase $\phi_\beta$
and $\phi_\gamma$. We suppose that 
$a_{\beta} = |a_{\beta}|^{1/2} exp(i\phi_\beta)$
and 
$a_{\gamma} = |a_{\gamma}|^{1/2} exp(i\phi_\gamma)$
and that these phases 
are random and uncorrelated. Thus if an average 
over many scattering experiments is taken we have 
$< e^{i\phi_\beta} e^{-i\phi_\gamma} > = \delta_{\beta\gamma} $
and consequently we find that 
$<|\Psi (y)|^{2}>  = |a_{\beta}|^{2} |\Psi_{\beta}|^{2}
+ |a_{\gamma}|^{2} |\Psi_{\gamma}|^{2}$ is the sum 
of scattering events in the different quantum channels. 
However, as soon 
as we are interested in quantities which depend to forth (or higher order)
on the amplitudes then even after averaging over the random phases 
of the incident waves we find that scattering processes 
which involve two (or more) incident waves matter. For instance 
consider $<|\Psi (y)|^{2} |\Psi (y^{\prime})|^{2}>
- <|\Psi (y)|^{2} > <|\Psi (y^{\prime})|^{2}>$ 
which is determined by 
$|a_{\beta}|^{2}|a_{\gamma}|^{2}
\Psi_{\beta}^{*}(y) \Psi_{\gamma}(y)
\Psi_{\beta}(y^{\prime})\Psi_{\gamma}^{*}(y^{\prime}) + h. c.$. 
This expression describes a correlation of the particle 
density at the points $y$ and $y^{\prime}$. Obviously 
for such higher order correlations superpositions are very important
even if we associate a random phase with each incident scattering states 
and take an average over these phases. 

We can relate such density fluctuations 
to functional derivatives of scattering matrix expressions. 
To describe density fluctuations of 
scattering processes with an incident 
carrier stream from channel $\beta$ and $\gamma$ 
we consider 
\begin{equation}
\nu(\alpha,y,\beta, \gamma) = -\frac{1}{4\pi i}
\left( S_{\alpha \beta}^{\dagger}
\frac{\delta S_{\alpha \gamma}}{\delta V(y)} - 
\frac{\delta S_{\alpha \beta}^{\dagger}}{\delta V(y)}
S_{\alpha \gamma}\right)\;\;
\label{fdos1}
\end{equation}
Note that two different scattering matrices enter into this 
expression. 
The connection to the scattering states is determined 
by the relation 
\begin{equation}
\sum_{\alpha} \nu(\alpha,y,\beta, \gamma) 
= (1/h) (v_{\beta} v_{\gamma})^{-1/2} 
\Psi_{\beta}^{*}(y) \Psi_{\gamma}(y)\;\;
\label{fdos2}
\end{equation}
Here $v_{\beta}$ and $v_{\gamma}$ are the (asymptotic) velocities of 
the carriers in the scattering channels ${\beta}$ and ${\gamma}$
far away from the scattering region. Eq. (\ref{fdos2}) was 
given in Ref. \cite{math} and a detailed derivation of this 
relation is presented in \cite{ankara}. 

The expressions $\nu(\alpha,y,\beta, \gamma)$ can be viewed 
as the off-diagonal elements of a partial density of states matrix.
In Eq. (\ref{fdos2}) we take the sum over out-going channels.
The resulting matrix is the {\it local} density of states matrix.  
Using Eq. (\ref{fdos1}) in Eq. (\ref{fdos2}) 
we find, 
\begin{equation}
\nu(y,\beta, \gamma)  \equiv \sum_{\alpha} \nu(\alpha,y,\beta, \gamma) 
= -\frac{1}{2\pi i} \sum_{\alpha}
\left( S_{\alpha \beta}^{\dagger}
\frac{\delta S_{\alpha \gamma}}{\delta V(y)} \right)\;\;
\label{fdos3}
\end{equation}
where we have taken into account that the scattering matrix is unitary. 

Let us now consider the total density of states matrix. 
The fluctuations 
of interest are then the total particle number fluctuations 
in the scattering region ${\Omega}$, 
$D(\beta, \gamma) = \int_{\Omega} dy \nu(y,\beta, \gamma)$. 
Furthermore, if the volume of integration is sufficiently large, 
we can, in WKB approximation, 
replace the functional derivative with respect 
to $V$ with a derivative with respect to energy.
The matrix which governs the fluctuations in the particle 
number in the scattering region then becomes,  
\begin{equation}
D(\beta, \gamma)
= \frac{1}{2\pi i} \sum_{\alpha}
\left( S_{\alpha \beta}^{\dagger}
\frac{dS_{\alpha \gamma}}{dE} \right)\;\;
\label{fdos4}
\end{equation}
Eq. (\ref{fdos4}) is the Wigner-Smith delay time matrix \cite{Smith}. 
We have earlier emphasized that the appearance of the energy derivative 
is a consequence of approximations (here the fact that we consider the 
WKB limit). We also mention that strictly speaking, here we do not 
consider a "delay". We do not compare 
with a reference scattering problem (a free motion) as is typically 
done in nuclear scattering problems. Eq. (\ref{fdos4}) determines 
a total time or absolute time and we should more appropriately 
call it the absolute time matrix instead of the delay time matrix.

The partial density of states matrix has 
been used in Ref. \cite{math} to obtain the second order 
in frequency term of the ac-conductance (see Eq. (\ref{ac1})). 
Ref. \cite{plb} investigated the current induced into a nearby 
gate due to charge fluctuations in quantum point contacts and 
chaotic cavities. More recently, the charge fluctuations in 
two nearby mesoscopic conductors was treated with this approach
and the effect of quantum dephasing due to charge fluctuations 
was calculated within this approach \cite{mbam,ankara}. 
The results can be compared with other theoretical works \cite{levinson}
and with experiments \cite{buks1}.

The Wigner-Smith delay time matrix has received wider attention. 
In recent years the focus has been on the calculation of the entire 
distribution function 
of delay times \cite{Sommers} for structures whose dynamic 
is in the classical limit chaotic (chaotic cavities). 
Predominantly structures have been investigated in  which 
carrier propagation is an allowed energy region. 

We conclude by briefly discussing the Wigner-Smith matrix 
for a tunnel barrier. For a symmetric barrier with 
transmission and reflection probability $T$ and $R$
the scattering matrix 
has elements 
$S_{11} = S_{22} = -i \sqrt{R} \exp(i\phi)$, 
and 
$S_{21} = S_{12} =\sqrt{T} \exp(i\phi)$
where $\phi$ is the phase accumulated during a reflection or 
transmission process. Thus the elements of the Wigner-Smith delay 
time matrix Eq. (\ref{fdos4})
are 
\begin{equation}
    {\cal D}_{11} = {\cal D}_{22} = 
    \frac{1}{2\pi} \frac{d\phi_{n}}{dE} ,\,\, 
    {\cal D}_{12} = {\cal D}_{21} = \frac{1}{4\pi}  
    \frac{1}{\sqrt{R_{n}T_{n}}} \frac{dT_{n}}{dE} .
\label{mqpc}
\end{equation}
To be specific consider now the case of a tunnel barrier. 
In the WKB limit we have $R \approx 1$ and 
$T = exp(-2S/\hbar)$ with 
$S = \int dy \sqrt{2m} \sqrt{V(y)-E}$
where the integral extends from one turning point to the other. 
We have $d\phi/dE = 0$ and 
using the expression $\tau_T = m \int dy (\sqrt{2m} \sqrt{V(y)-E})^{-1}$
for the traversal time of tunneling gives 
$dT/dE = (2\tau_T/\hbar) T$. Consequently, the diagonal 
elements of the Wigner-Smith matrix vanish and the non-diagonal elements 
are 
\begin{equation}
    {\cal D}_{12} = {\cal D}_{21} = \frac{\tau_T}{2\pi\hbar} \sqrt T .
\label{mqpc1}
\end{equation}
Thus while the the average density inside the barrier vanishes 
(the trace of the Wigner-Smith matrix is zero)
the off-diagonal elements are non-zero and indicate that the fluctuations 
of the charge will in general be non-vanishing even deep inside 
the classically forbidden region. 

The discussion of this section rests admittedly vague. 
While important first steps have been made to extend the notion 
of partial density of states to treat fluctuations, 
our discussion shows that 
even at the conceptual level, there is clearly room for more research. 

Another development which could be discussed here is a theory of 
quantum pumping in small systems. 
In quantum pumping one is interested in the current generated as 
two parameters (like gate voltages, magnetic fluxes) 
which modulate the system are varyed sinusoidally but out of phase. 
Brouwer \cite{BR1}, Avron et al. \cite{Avron}, Shutenko et al. \cite{alein}
and Polianski and Brouwer \cite{BR2} develop a theory 
which is based on the modulation of the partial densities of states 
discussed here.

\section{Discussion}

The Larmor clock and its close relatives have become one of the most widely 
investigated approaches mainly in order to understand the question: "How 
long does a particle traveling through a classically forbidden region 
(a tunnel barrier) interact with this region?". We have already in the 
previous sections pointed out that there is no consensus in the 
interpretation even of this simple clock. 
Regardless of these difficulties the investigation of the 
Larmor clock has been helpful in understanding a number of transport problems:
In particular we have  discussed a hierarchy of density of states as they 
occur in open multiprobe mesoscopic conductors. These density of states
are directly related to local Larmor times. We have shown that a small
absorption or a small emission of particles can be described with these 
densities (or in terms of the Larmor times). We have shown that the 
transmission probabilities through weakly coupled contacts like the 
STM is related to these densities. We have shown that a weakly
coupled voltage probe, describing inelastic scattering or a dephasing
process can be treated in terms of these densities. We have also pointed 
out that the ac-conductance of a mesoscopic conductor at small frequencies
can be expressed with the help of these densities. Furthermore, we have 
indicated that it is useful to consider also the off-diagonal elements of 
a partial density of states matrix since this permits a description 
of fluctuation processes.  
Thus there is no 
question that the investigation of the Larmor clock has been a very 
fruitful and important enterprise.


\begin{thebibliography}{9.}
\addcontentsline{toc}{section}{References}
%
\bibitem{BAZ}     A. I. Baz', Sov. J. Nuc. Phys. {\bf 4}, 182 (1967);
                  {\bf 5}, 161 (1967). 

\bibitem{RYB}     V. F. Rybachenko, Sov. J. Nucl. Phys. {\bf 5}, 635 (1967). 

\bibitem{LARMOR}  M. B\"uttiker, Phys. Rev. B {\bf 27}, 6178 (1983).

\bibitem{LEAE1}   C. R. Leavens and G. C. Aers, 
                  Solid State Commun. {\bf 63}, 1107 (1989). 

\bibitem{LEAE2}   C. R. Leavens and G. C. Aers
                  Phys. Rev. B {\bf 40}, 5387-5400 (1989). 

\bibitem{BU1}     M. B\"{u}ttiker,
                  J. Phys.: Condensed Matter {\bf 5}, 9361 (1993).
                  
                  

\bibitem{BTP}     M. B\"{u}ttiker, H. Thomas, and A. Pr\^etre,
                  Z. Phys. B {\bf 94}, 133 (1994). 
                  
                
\bibitem{BTP1}    M. B\"{u}ttiker, A. Pr\^{e}tre and H. Thomas,
                  Phys. Rev. Lett. {\bf 70}, 4114 (1993); 
                  M. B\"{u}ttiker, H. Thomas, and A. Pr\^{e}tre,
                  Phys. Lett. A {\bf 180}, 364 (1993).
                      
\bibitem{GASP}    V. Gasparian, T. Christen, and M. B\"{u}ttiker, 
                  Phys. Rev. A {\bf 54}, 4022 (1996). 

\bibitem{GRAM}    T. Gramespacher and M. Buttiker, 
                  Phys. Rev. B {\bf 56}, 13026-13034 (1997);
                  Phys. Rev. B {\bf 60},  2375-2390 (1999); 
                  Phys. Rev. B {\bf 61}, 8125-8132 (2000).                               

\bibitem{dash}    R. Dashen, S. Ma, and H. J. Bernstein, 
                  Phys. Rev. {\bf 187}, 345 (1969). 

\bibitem{avba}    Y. Avishai and Y. B. Band, 
                  Phys. Rev. B {\bf 32}, 2674 (1985).                

                   
\bibitem{ianna1}  G. Iannaccone, Phys. Rev. B {\bf 51}, 4727 (1995).  

\bibitem{ianna2}  G. Iannaccone and B. Pellegrini, 
                  Phys. Rev. B {\bf 53}, 2020 (1996).  
                  
                 
\bibitem{Land}    M. B\"{u}ttiker and R. Landauer, Phys. Rev. Lett.
                  {\bf 49}, 1739 (1982); Physica  Scripta {\bf 32},
                  429-434, (1985).                  
                  
                  
\bibitem{SOKOL}   D. Sokolovski and L. M. Baskin
                  Phys. Rev. A {\bf 36}, 4604-4611 (1987); 
                  H. A. Fertig, Phys. Rev. B {\bf 47}, 1346-1358 (1993).

\bibitem{GOR}     V. Gasparian, M. Ortuno,  J. Ruiz, and E. Cuevas 
                  Phys. Rev. Lett. {\bf 75}, 2312 (1995); Y. Japha 
                  and G. Kurizki, Phys. Rev. A {\bf 60}, 1811 (1999). 
                 
\bibitem{Stein}   A. M. Steinberg, Phys. Rev. Lett. {\bf 74}, 2405 (1995).  

\bibitem{muga}    S. Brouard, R. Sala,
                  J. G. Muga, Phys. Rev. A {\bf49} 4312 (1994).                

\bibitem{zhao}    X. Zhao, J. Phys. Cond. Matter, {\bf 12}, 4053 (2000).
                  
\bibitem{MB90}    M. B\"{u}ttiker, in "Electronic Properties of Multilayers and low
                  Dimensional Semiconductors",
                  edited by J. M. Chamberlain, L. Eaves, and J. C. Portal,
                  (Plenum, New York, 1990). p. 297-315.
                  
       
\bibitem{rama}    S. A. Ramakrishna and N. Kumar, 
                  Phys. Rev. B {\bf 61}, 3163 (2000).     

\bibitem{been}    C.W.J. Beenakker, cond-mat/0009061                  

\bibitem{br2}     P. W. Brouwer, S. A. van Langen, K. M. Frahm,
                  M. B\"uttiker, and C. W. J. Beenakker, 
                  Phys. Rev. Lett. {\bf 79}, 
                  914 (1997).
                  
\bibitem{STM}     G. Binnig and H. Rohrer, 
                  Helv. Phys. Acta {\bf 55}, 726 (1982); 
                  J. Tersoff and D. R. Hamann, 
                  Phys. Rev. B {\bf 31}, 805 (1985). 
                   
\bibitem{MB88}    M. B\"{u}ttiker,
                  IBM J. Res. Develop. {\bf 32}, 63 (1988).

\bibitem{BB}      P. W. Brouwer and C. W. J. Beenakker, 
                  Phys. Rev. B {\bf 55}, 4695 (1997). 
                                                     

\bibitem{JAY}     T. P. Pareek, Sandeep K. Joshi, A. M. Jayannavar, 
                  Phys. Rev. B {\bf57}, 8809 (1998). 
                    
   
\bibitem{MBOPT}   M. B\"{u}ttiker,
                  in "Analogies in Optics and Micro-Electronics",
                  edited by W. van Haeringen and D. Lenstra,
                  Kluwer Academic Publishers,
                  (Dordrecht-Boston-London, 1990). p. 185-202.

\bibitem{TCMB}    T. Christen and M. B\"{u}ttiker, 
                  Phys. Rev. Lett. {\bf 77}, 143 (1996).
         
\bibitem{Wees}    B. J. van Wees et al.,
                  Phys. Rev. Lett. {\bf 60}, 848 (1988);
                  D. A. Wharam et al.,
                  J. Phys. C: Solid State Phys. {\bf 21}, L209
                  (1988).
                  
\bibitem{CURACAO}  M. B\"{u}ttiker and T. Christen, in 
                   "Mesoscopic Electron Transport", 
                   NATO Advanced Study Institute, Series E: Applied Science, 
                   edited by L. L.  Sohn, L. P. Kouwenhoven and G. Schoen,
                   (Kluwer Academic Publishers, Dordrecht, 1997). 
                   Vol. 345. p. 259. 
                   cond-mat/9610025  
                   
                  
\bibitem{Guo}      Tiago De Jesus, Hong Guo, and Jian Wang, 
                   Phys. Rev. B {\bf 62}, 10774 (2000).  

\bibitem{pp}       J. P. Pieper and and J. C. Price, 
                   Phys. Rev. Lett. {\bf 72}, 3586 (1994). 

\bibitem{desrat}   W. Desrat, D. K. Maude, L. B. Rigal, M. Potemski, 
                   J. C. Portal,
                   L. Eaves, M. Henini,  Z. R. Wasilewski, A. Toropov, 
                   G. Hill and M. A. Pate,   
                   Phys. Rev. B {\bf 62} 12990 (2000).  

      
                              

\bibitem{Mikh}     S. A. Mikhailov and V. A. Volkov, JETP Lett. {\bf 61}, 
                   524 (1995). 
\bibitem{math}
                   M. B\"uttiker, J. Math. Phys., {\bf 37}, 4793 (1996).              

\bibitem{ankara}   M. B\"{u}ttiker, 
                   in "Quantum Mesoscopic Phenomena and Mesoscopic Devices", 
                   edited by I. O. Kulik and R. Ellialtioglu, (Kluwer, 
                   Academic Publishers, Dordrecht, 2000). Vol. 559, p. 211.
                   cond-mat/9911188                 
\bibitem{Smith}
                   F. T. Smith, Phys.\ Rev.\ {\bf 118} 349 (1960).  


                                     
\bibitem{plb}      M. H. Pedersen, S. A. van Langen and M. B\"{u}ttiker,
                   Phys. Rev. B {\bf 57}, 1838 (1998). 
\bibitem{mbam}     
                   M. B\"uttiker and A. M. Martin, 
                   Phys. Rev. B {\bf 61}, 2737 (2000).
                   
   
\bibitem{levinson} 
                   Y. B. Levinson, Europhys. Lett. {\bf 39}, 299 (1997);
                   L. Stodolsky, Phys. Lett. B {\bf 459}, 193 (1999).                                                            


\bibitem{buks1}                   
                   E. Buks, R. Schuster, M. Heiblum, D. Mahalu and V. Umansky,
                   Nature {\bf 391}, 871 (1998);                    
                   D. Sprinzak, E. Buks, M. Heiblum and H. Shtrikman,
                   Phys. Rev. Lett. {\bf 84}, 5820 (2000). 
                                                                          
\bibitem{Sommers}  Y.~V.\ Fyodorov and
                   H.~J.\ Sommers, Phys. Rev. Lett. {\bf 76}, 4709 (1996);
                   V.~A.\ Gopar, P.~A.\ Mello, and M.\
                   B\"{u}ttiker, Phys.\ Rev.\ Lett.\ {\bf 77}, 3005 (1996);
                   P.~W.\ Brouwer, K.~M.\ Frahm, and C.~W.~J.\ Beenakker,
                   Phys.\ Rev.\ Lett.\ {\bf 78}, 4737 (1997);
                   C. Texier and A. Comtet, {\it ibid.}{\bf 82}, 4220 (1999). 
                   
\bibitem{BR1}      P.W. Brouwer, Phys. Rev. B 58, R10 135 (1998).                   

\bibitem{Avron}    J. E. Avron, A. Elgart, G. M. Graf, and L. Sadun, 
                   Phys. Rev. B {\bf 62}, R10618 (2000). 

\bibitem{alein}    T. A. Shutenko, I. L. Aleiner, B. L. Altshuler, 
                   Phys. Rev. B{\bf 61}, 10366 (2000).                    

\bibitem{BR2}      M. L. Polianski, P. W. Brouwer, cond-mat/0102159 
                                                      





                  
\end{thebibliography}
\end{document}